\documentclass[twocolumn,showpacs,superscriptaddress,preprintnumbers,amsmath,amssymb]{revtex4-2}
\usepackage{graphicx}
\usepackage{dcolumn}
\usepackage{appendix}
\usepackage{bm}

\usepackage[usenames,dvipsnames]{xcolor}
\usepackage{subfigure}
\usepackage{bbm}
\usepackage{enumerate}
\usepackage[
  bookmarks=true,
  colorlinks,
  linkcolor=blue,
  urlcolor=blue,
  citecolor=blue,
  plainpages=false,
  pdfpagelabels,
  final,
  breaklinks=true
]{hyperref}

\usepackage{titlesec}
\usepackage{array}
\usepackage{dsfont}
\usepackage{physics}
\usepackage{mathpazo}

\begin{document}

\title{Quantum stochastic analysis of non-linear driven light emission}

\author{Philipp Stammer}
\email{philipp.stammer@icfo.eu}
\affiliation{ICFO-Institut de Ciencies Fotoniques, The Barcelona Institute of Science and Technology, Castelldefels (Barcelona) 08860, Spain.}
\affiliation{Atominstitut, Technische Universit\"{a}t Wien, 1020 Vienna, Austria}

\date{\today}

\begin{abstract}

This work extends quantum optical models of high harmonic generation by considering a quantum stochastic analysis of the field modes coupled to an environment.
In particular, we study the open system dynamics by solving the quantum Langevin equation for a non-linear driven cavity coupled to the environment.
For an unstructured environment without memory, we show that the emission characteristics of the intense driven cavity is isomorphic to the process of high harmonic generation and a non-linear antenna. This is achieved by using the quantum regression theorem for the Markovian dynamics, and further allows to obtain the upper bound of the radiated power due to the emitter fluctuations.
This opens the path to connect current quantum optical approaches of HHG with open system descriptions and towards driven-dissipative systems in the non-linear regime.

\end{abstract}

\maketitle

Photon counting from a resonator is a probabilistic process, contrary to the oscillations of a classical Hertzian dipole which is a deterministic system. Nevertheless, in their emission characteristic the antenna and the driven resonator are isomorphic~\cite{milonni2010laser}. 
To be precise, the emission of a Hertzian dipole (i.e. the Poynting theorem), and the driven resonator (i.e. the Wiener-Khinchin theorem of a stationary random process) are isomorphic for the observer perspective of the power spectrum. 
In the following, we show that a similar isomorphism holds for the process of high-order harmonic generation~\cite{lewenstein1994theory} and a non-linear driven leaky cavity. This is achieved by a quantum stochastic analysis of the underlying dynamics.

Quantum stochastic descriptions of physical processes are a powerful tool to capture the inherently probabilistic and stochastic nature of quantum phenomena.
Stochastic quantum dynamics reveal important aspects in quantum optics~\cite{gardiner2004quantum}, quantum information theory~\cite{wiseman2009quantum, nielsen2010quantum} or quantum thermodynamics~\cite{strasberg2022quantum}. 
It is particularly important for the investigation of open quantum systems~\cite{breuer2002theory}, in which stochastic noise from an environment influences the system dynamics. 
Yet, for the non-linear driven process of high harmonic generation (HHG) it has thus far remained a disconnected concept, despite the recent progress in the quantum optical formulation of HHG~\cite{cruz2024quantum}.

The discussion in this Letter is complementary to the isomorphism between a cavity and an antenna, and extends the recent advances in the quantum optical description of the process of high-order harmonic generation~\cite{cruz2024quantum, stammer2025theory, stammer2023quantum}. 
First, we extend the isomorphism of a driven cavity to the non-linear regime connecting with HHG, and in addition open the discussion of quantum HHG towards open systems and quantum noise~\cite{gardiner2004quantum}. 
So far, quantum HHG was mainly focused on describing the quantum state of the field after the HHG process~\cite{gorlach2020quantum, lewenstein2021generation}. It was shown that HHG allows to generate entangled states~\cite{stammer2024entanglement, theidel2024evidence, stammer2022theory, yi2024generation}, quadrature squeezing~\cite{stammer2024entanglement, lange2024electron, tzur2024generation, yi2024generation, rivera2024squeezed, lange2025excitonic}, and that post-selection techniques allow for the reconstruction of optical cat states~\cite{lewenstein2021generation, stammer2022high, rivera2022strong}. 
With this work, we extend the discussion on the quantum optical formulation of HHG to the realm of cavity QED and quantum stochastic dynamics. 
This is achieved by using the approach of the quantum Langevin equation for the evolution of the field operators, naturally incorporating the environmental noise operators and can facilitate treatments beyond the Markov approximation. 

In the following, we first show the isomorphism between the nonlinear driven cavity and the process of HHG. Considering an unstructured environment without memory effects allows to use the quantum regression theorem to solve the dynamics, introducing the notion of open quantum systems to strong field driven dynamics and quantum optical HHG. Using these results we then provide an upper bound for the radiated power due to the fluctuations of the non-linear emitter.

\begin{figure}
    \centering
	\includegraphics[width = 0.9\columnwidth]{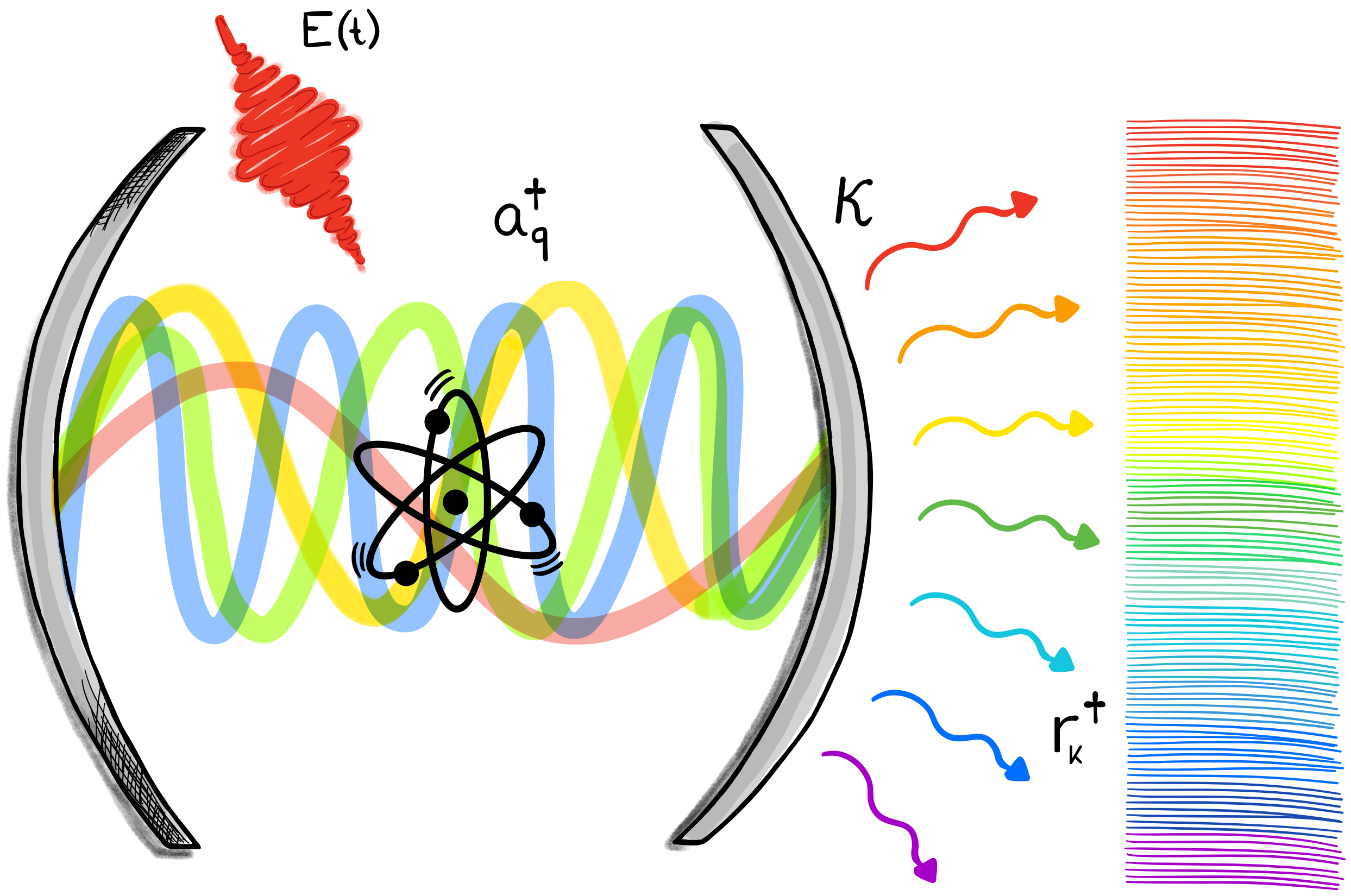}
	\caption{Cartoon of the non-linear driven cavity, with the emitter driven by an intense coherent field $E(t)$ leading to emission into the higher-order cavity modes $a_q^\dagger$. Leakage of the cavity field into the continuous reservoir modes $r_k^\dagger$ with rate $\kappa$ (in the Markovian regime), allows to show the isomorphism to the HHG process and the non-linear antenna.   
    }
      \label{fig:scheme}
\end{figure}

\section*{\label{sec:1}Quantum Langevin equation}

The system of interest is a non-linear driven emitter in a cavity coupled to the radiation field of the higher-order resonator modes. We consider the classical driving field to be in a coherent state $\ket{\alpha}$, and the matter system is initially in its ground state $\ket{g}$. The corresponding Hamiltonian of this system is given by~\cite{stammer2025theory};
\begin{equation}
\label{eq:hamiltonian_total}
    H = \sum_{q=1}^{q_c} \omega_{q} a_{q}^\dagger a_{q} + H_S + H_{I}, 
\end{equation}
with generic emitter Hamiltonian $H_S$, and the higher-order resonator modes $q \ge 2$ of the fundamental cavity mode $q=1$.
The interaction Hamiltonian $H_I$ couples the dipole $d$ of the emitter to the resonator 
\begin{equation}
\label{eq:hamiltonian_interaction}
    H_{I} = - \, d \, E_Q, 
\end{equation}
with the electric field operator given by 
\begin{equation}
    E_Q = - i \sum_{q=1}^{q_c} g_q \left( a_{q}^\dagger - a_{q} \right),
\end{equation}
where $g_q$ is the coupling constant of the light-matter interaction and the operators fulfill the standard bosonic commutation relation $[a_q, a_p^\dagger] = \delta_{qp}$.
For the emitter dynamics, we only consider the intense classical driving field and neglect the quantum backaction from the field fluctuations~\footnote{Due to the high intensity of the driving field, necessary for inducing non-linear dynamics in the emitter, the backaction of the change in the field on the atom can be neglected~\cite{diestler2008harmonic}. Effectively this leads to the assumption that the emitter interacts with a classical driving field of constant amplitude.}. This allows to propagate the dipole via the semi-classical interaction, consistent with the strong field picture of the intense driving field inducing non-linear dynamics of the emitter~\cite{stammer2025theory, diestler2008harmonic}.

To analyze the influence of an environment, we consider the coupling of the cavity modes to a continuum of environmental modes 
\begin{equation}
    H_E = \int dk \left[  \omega_k r_k^\dagger r_k + \sum_{q} \left( g_k a_q^\dagger r_k + g_k^* a_q r_k^\dagger \right) \right],
\end{equation}
where $r_k^{(\dagger)}$ are the annihilation (creation) operators of the environment, and $g_k$ are the resonator-environment couplings. 
The Heisenberg-Langevin equation of motion (EOM) for the cavity mode is accordingly given by 
\begin{equation}
\label{eq:cavity_EOM}
    \dv{t} a_q(t) = - i \omega_q a_q(t) + g_q \, d(t) - i \int dk \ g_k r_k(t),
\end{equation}
and the environment operator follows the EOM 
\begin{equation}
    \dv{t} r_k(t) = - i \omega_k r_k(t) - i g_k a_q(t), 
\end{equation}
with the formal solution 
\begin{equation}
    r_k(t) = e^{- i \omega_k t } \left[ r_k(0) - i g_k \int_0^t dt' \, a_q(t') e^{i \omega_k t'}  \right].
\end{equation}

We therefore have the quantum Langevin EOM for the cavity mode with a non-linear driven emitter coupled to an environment
\begin{align}
\label{eq:quantum_Langevin}
    \dv{t} a_q(t) = & - i \omega_q a_q(t) + g_q \, d(t)  - i \int dk \ g_k \ r_k(0) e^{- i \omega_k t} \nonumber \\
    & - \int dk \ g_k^2 \ e^{- i \omega_k t} \int_0^t dt' \, a_q(t') e^{i \omega_k t'} . 
\end{align}

The last two terms correspond to the quantum noise operator and the damping term with memory, respectively. 
If we assume that the environment into which the cavity is leaking has no structure (white noise), such that we have $g_k \approx g_0$ without any $k$ dependence, we can write 
\begin{align}
    \int dk \, e^{- i \omega_k (t-t')} = \frac{2\pi}{c} \delta (t-t'), 
\end{align}
where $c = \omega_k /k$ is the speed of light. Consequently, we have
\begin{align}
\label{eq:EOM_field}
    \dv{t} a_q(t) = & \left[ - i \omega_q - \kappa \right] a_q(t) + g_q \, d(t) \\
    & - ig_0 \int dk \ r_k(0) \, e^{- i \omega_k t} \nonumber,
\end{align}
where $\kappa = g_0^2\pi /c$ is the cavity damping~\footnote{Note that there is an additional factor of $1/2$ due to preserving the commutator relation when evaluating the delta function in the time-integral. From $[a_q(t),a_q^\dagger(t)] =1$ it follows that $ \int_0^t dt' f(t') \delta(t-t') = \frac{1}{2} f(t)$.}. 
The first and third term correspond to the coherent dynamics of the mode, whereas the second and fourth term are the quantum noise due to the coupling with the environment. Originating from the assumption of the unstructured environment we find that the last term of Eq.~\eqref{eq:quantum_Langevin} has no memory effect since the mode operator only depends on the time $t$, and not the entire history, giving rise to the damping with constant rate $\kappa$. 
Here, we note that this approximation of a frequency independent coupling constant is considered as the \textit{first Markov approximation}~\cite{gardiner1985input}. Due to this, memory effects are absent, which will later become important since it allows to use the quantum regression theorem for the derivation of two-time correlation functions~\cite{lax1963formal}. 
We can now write the solution of the quantum Langevin equation for the cavity operator
\begin{align}
    a_q(t)  = & \ e^{- (i \omega_q + \kappa) t} a_q(0) + \int dk \ r_k(0) B_k(t)  \\
    & + g_q \int_0^t dt^\prime e^{-(i\omega_q + \kappa) (t-t^\prime)} d(t^\prime), \nonumber
\end{align}
where we have defined 
\begin{align}
    B_k(t) = - i g_0 \int_0^t dt^\prime \, e^{ -(i \omega_q + \kappa) (t- t^\prime)} e^{- i \omega_k t'}.
\end{align}

Before computing expectation values of the field, we first have a closer look at the non-linear driven emitter. We therefore decompose the time-dependent dipole moment $d(t)$ into its mean and the fluctuations
\begin{align}
    d(t) = \expval{d(t)} + \delta d(t),
\end{align}
where $\expval{d(t)} = \bra{g} d(t) \ket{g}$ and $\delta d(t) = d(t) - \expval{d(t)}$ are the mean value of the dipole and its fluctuation, respectively. Note that all expectation values are taken with respect to the initial state $\ket{g}$, which is given by the emitter ground state.  
While the term corresponding to $\expval{d(t)}$ represents a coherent driving input, the $\delta d(t)$ contribution acts as an additional noise term to the field quantities. 
Using this, we have the dipole noise operator 
\begin{align}
    \mathfrak{D}(t) = g_q \int_0^t dt^\prime e^{-(i\omega_q + \kappa) (t-t^\prime)} \delta d(t^\prime),
\end{align}
such that 
\begin{align}
    a_q(t)  =& e^{- (i \omega_q + \kappa) t} a_q(0) + g_q \int_0^t dt^\prime e^{-(i\omega_q + \kappa) (t-t^\prime)} \expval*{d(t^\prime)}  \nonumber  \\
    & + \int dk \, r_k(0) B_k(t) + \mathfrak{D}(t).
\end{align}

This is the final quantum Langevin equation for a non-linear driven emitter in a resonator coupled to the environment. This dissipative non-linear emitter model is within the Markovian approximation for an unstructured environment~\footnote{The non-Markovian case for a structured environment, showing the influence of the environment on the spectral features of the non-linear emission, will be discussed elsewhere.}. 
Now, we shall first look at the average photon number occupation of the cavity modes
\begin{align}
\label{eq:average_occupation}
    \expval*{a_q^\dagger(t) a_q(t)} = & \ g_q^2 \ e^{- 2 \kappa t} \abs{\int_0^t dt' e^{(i \omega_q + \kappa) t'} \expval{d(t')} }^2 \\
    & + \expval*{\mathfrak{D}^\dagger(t) \mathfrak{D}(t)}, \nonumber 
\end{align}
where we have used that all field modes (cavity and environment) are initially in the vacuum~\footnote{For the optical frequencies considered here, the assumption of a vacuum environment is a good approximation considering that the photon energy in the optical regime is on the order of eV.}, i.e. $\ket{\psi(0)} = \ket{\{ 0_q \}, \{ 0_k\}, g }$.
We identity the first term as the coherent contribution to the occupation number, while the second term results from the emitter fluctuations~\cite{stammer2025theory}. 
To gain more insights into the influence of the non-linearity of the process, we consider a Fourier decomposition of the time-dependent dipole moment~\cite{lewenstein1994theory};
\begin{align}
    \expval{d(t)} = \sum_{N=-\infty}^\infty d(\omega_N) e^{- i \omega_N t}, 
\end{align}
with Fourier coefficients $d(\omega_N)$ of the dipole expectation value. We can therefore express the time integral in \eqref{eq:average_occupation} as
\begin{align}
    \int_0^t dt' e^{(i \omega_q + \kappa )t'} \expval{d(t')} = \sum_{N=0}^\infty d(\omega_N) \frac{e^{i(\omega_q - \omega_N) t + \kappa t} - 1 }{i (\omega_q - \omega_N) + \kappa},
\end{align}
where we have only kept the positive frequency components due to a \textit{rotating-wave like approximation}~\cite{gardiner1985input}. 
With this in hand, we can evaluate the absolute value in Eq.~\eqref{eq:average_occupation}, for which we find 
\begin{align}
\label{eq:double_sum}
    \sum_{N} & \left[ \abs{d(\omega_N)}^2 \abs{\frac{e^{i(\omega_q - \omega_N) t + \kappa t} - 1 }{i (\omega_q - \omega_N) + \kappa}}^2 \right. \\
    & \left. + d(\omega_N) \sum_{M \neq N} d^*(\omega_M) f(t;\omega_N, \omega_M) \right], \nonumber
\end{align}
where $f(t; \omega_N, \omega_M)$ is a function of the frequencies $\omega_N$ and $\omega_M$~\footnote{The explicit expression for the function $f$ in Eq.~\eqref{eq:double_sum} is given by $f(t;\omega_N, \omega_M) = [(e^{i(\omega_q-\omega_N)t + \kappa t} -1 )/(i(\omega_q - \omega_N) + \kappa)] [(e^{-i(\omega_q-\omega_M)t + \kappa t} -1 )/(-i(\omega_q - \omega_M) + \kappa)] $. }. 
For the first term in \eqref{eq:double_sum} we obtain 
\begin{align}
    \sum_N \frac{\abs{d(\omega_N)}^2}{(\omega_q - \omega_N)^2 + \kappa^2} \left[ 1 + e^{2 \kappa t} - 2 e^{\kappa t} \cos[(\omega_q - \omega_N)t] \right], 
\end{align}
and hence, for the long time limit of the average photon number we find 
\begin{align}
\label{eq:long_time_photon_number}
    \lim_{t \to \infty} \expval*{a_q^\dagger(t)a_q(t)} = g_q^2 \left( \ \sum_N \frac{\abs{d(\omega_N)}^2}{(\omega_q - \omega_N)^2 + \kappa^2} +  \frac{ \Delta_{\delta d}}{2 \kappa} \right),
\end{align}
where we have neglected the second term in \eqref{eq:double_sum} in which $M \neq N$ due to the fast oscillations. We further added the constant offset from the dipole noise term of Eq.~\eqref{eq:average_occupation}, which is of delta-correlated white noise type (see Appendix~\ref{app:dipole_noise}). 
This expression shows the equilibrium occupation between the energy gain due to the driving of the non-linear emitter, and the loss from the leaky cavity into the environment. 

We can now proceed and compute the power spectrum of the process, which is related to the first order auto-correlation function due to the Wiener-Khinchin theorem~\cite{wiener1930generalized, khintchine1934korrelationstheorie};
\begin{align}
\label{eq:WKT}
    S(\omega) = \frac{1}{\pi} \operatorname{Re}\left[ \int_0^\infty d\tau \lim_{t \to \infty} \expval*{a_q^\dagger (t) a_q (t+\tau)} e^{i \omega \tau} \right]. 
\end{align}

First, we need to find the two-time correlation function of the cavity mode, which we shall obtain by making use of the Onsager-Lax quantum regression theorem~\cite{lax1963formal, carmichael2013statistical}, which holds under the assumption of an unstructured, and therefore memory-less Markovian environment~\cite{ford1996there, talkner1986failure}. 
The quantum regression theorem (QRT), formulated by Lax~\cite{lax1963formal}, allows to compute multi-time correlation functions from single-time expectation values, in analogy to the classical regression theorem for stochastic processes introduced by Onsager~\cite{onsager1931reciprocal, onsager1931reciprocalII}. 
We can now use the QRT, and that we know how the field operator commutes with the Hamiltonian from Eq.~\eqref{eq:cavity_EOM}, such that we have (see Appendix \ref{app:regression_theorem})
\begin{align}
    \dv{\tau} & \expval*{a_q^\dagger (t) a_q(t+\tau)} =  [- i \omega_q - \kappa] \expval*{a_q^\dagger (t) a_q(t+\tau)} \nonumber \\
    & + g_q^2 \int_0^t dt^\prime e^{(i \omega_q-\kappa)(t-t^\prime)} \expval*{d(t^\prime)d(t+\tau)}.
\end{align}

The solution is readily found and given by 
\begin{align}
    \langle  a_q^\dagger & (t) a_q(t+\tau)\rangle =  e^{- (i \omega_q + \kappa) \tau} \expval*{a_q^\dagger (t) a_q(t)} \nonumber \\
    & + g_q \expval*{a_q^\dagger(t)} \int_0^\tau d\tau^\prime e^{- (i \omega_q + \kappa)(\tau - \tau^\prime)} \expval*{d(t+\tau^\prime)} \\
    & + \Delta_{\delta d} \frac{ g_q^2 }{2 \kappa} e^{- i \omega_q \tau} e^{- \kappa \tau} \left( 1- e^{- 2 \kappa t} \right). \nonumber
\end{align}

It remains to compute the average of the cavity mode
\begin{align}
\label{eq:field_average}
    \expval*{a_q^\dagger(t)} =  \sum_{N=0}^\infty A_N^* \left[ e^{i \omega_N t } - e^{i \omega_q t} e^{- \kappa t} \right],
\end{align}
where we have used that $d(-\omega_N) = d^*(\omega_N)$, and defined 
\begin{align}
    A_N^* = \frac{g_q \ d^*(\omega_N)}{- i (\omega_q - \omega_N) + \kappa} \, . 
\end{align}

In addition we have used that in the initial state the field is in the vacuum. 
In Eq.~\eqref{eq:field_average}, the first term is the coherent scattering from the non-linear driven "laser"-emitter with frequencies $\omega_N = N \omega$ of the anharmonic oscillating dipole, while the second term is the scattering of the cavity modes with resonator frequencies $\omega_q = q \omega$. The non-linear "laser"-emitter could, for instance, be the coherent emission of the HHG process~\cite{lewenstein2021generation, stammer2023quantum, stammer2024absence}.
As expected, the resonator scattering shows saturation due to $\kappa$, originating from the coupling to the environment. 
Considering long time scales, $\kappa t \gg 1$, we have
\begin{align}
    \expval*{a_q^\dagger(t)} \stackrel{\kappa t \gg 1}= \sum_{N=0}^\infty A_N^* \ e^{i \omega_N t}.
\end{align}

This shows that the coherently driven cavity, i.e. oscillator, oscillates with the perturbation $\omega_N$ and not with its own frequencies $\omega_q$. With this, we can now write
\begin{align}
    & \expval*{a_q^\dagger(t)a_q(t+\tau)} = e^{- (i \omega_q + \kappa)\tau} \left[ \expval*{N_q(t)}  + C_\Delta \left(1- e^{- 2\kappa t}\right) \right] \nonumber \\
    & + \expval*{a_q^\dagger(t)} \sum_{N=0}^\infty A_N e^{- i \omega_N t} \left[ e^{- i \omega_N \tau} - e^{- i \omega_q \tau} e^{- \kappa \tau} \right],
\end{align}
where $C_\Delta = \Delta_{\delta d} g_q^2 /(2 \kappa)$, and we have abbreviated the photon number operator $N_q(t) \equiv a_q^\dagger(t) a_q(t)$.
At this point, we emphasize that a factorization of the two-time average $\expval*{a_q^\dagger(t)a_q(t+\tau)}$, would imply that there are no temporal correlations~\footnote{Note that if we would only have one Fourier component of the coherent laser-emitter, e.g. for a linearly responding emitter with $N=1$, then the second term would read $ e^{- i\omega t}\expval*{a_q^\dagger(t)} \expval*{a_q(\tau)} $.}.
However, due to the non-linearity of the process this is not the case, and the correlations can manifest in squeezing, entanglement or anti-bunching properties~\cite{stammer2024entanglement, stammer2025theory}.
Now considering Eq.~\eqref{eq:long_time_photon_number}, we further have that 
\begin{align}
    \lim_{t \to \infty} \expval*{a_q^\dagger(t)a_q(t)} = \sum_{N=0}^\infty \abs{A_N}^2 + C_\Delta.
\end{align}

Consequently, in the long time limit, the field correlation function $\lim_{t \to \infty} \expval*{a_q^\dagger(t) a_q(t+\tau)} \equiv \expval*{a_q^\dagger(0) a_q(\tau)}_\infty$, is given by 
\begin{align}
    & \expval*{a_q^\dagger(0) a_q(\tau)}_\infty = e^{- i \omega_q \tau} e^{- \kappa \tau} \left( \sum_{N=0}^\infty \abs{A_N}^2 + 2 \,C_\Delta \right) \\
    & + \sum_{N,M=0}^\infty A_N^* A_M e^{i (\omega_N - \omega_M) t} \left[ e^{- i \omega_M \tau} - e^{- i \omega_q \tau} e^{- \kappa \tau} \right], \nonumber
\end{align}
and due to the fast oscillations whenever $N\neq M$, we only consider the terms in which $N=M$, such that 
\begin{align}
    \expval*{a_q^\dagger(0) a_q(\tau)}_\infty = \sum_{N=0}^\infty \abs{A_N}^2 e^{- i \omega_N \tau} + 2  \, C_\Delta \, e^{- (i \omega_q + \kappa) \tau}.
\end{align}

Interestingly, we find that in the long-time limit only the incoherent part from the dipole correlations ($C_\Delta$) is affected by $\kappa$, and the coherent scattering ($\abs{A_N}^2$) does not decay. This would allow to experimentally distinguish the two contributions in two-time expectation values.
From this, we can directly compute the power spectrum $S(\omega) = S_{coh}(\omega) + S_\Delta (\omega)$, composed of the elastic scattering
\begin{align}
\label{eq:spectrum_markov}
    S_{coh}(\omega) = g_q^2 \sum_{N=0}^\infty \frac{\abs{d(\omega_N)}^2}{(\omega_q - \omega_N)^2 + \kappa^2} \delta(\omega - \omega_N), 
\end{align}
and the contribution of the dipole fluctuations
\begin{align}
    S_{\Delta} (\omega) =  2\, C_\Delta  \frac{\kappa}{(\omega - \omega_q)^2 + \kappa^2}.
\end{align}
We can see that the power spectrum of the elastic scattering consists of a series of peaks at the Fourier frequencies $\omega_N$ of the dipole, where the height of each peak is determined by the strength of the respective Fourier component of the dipole moment.
In contrast, the incoherent contribution shows a continuous distribution~\cite{stammer2025theory}.
Looking at the emitted power, obtained by integrating over the power spectrum, we have
\begin{align}
    P = \int d \omega \, S(\omega) = g_q^2 \sum_{N=0}^\infty \frac{\abs{d(\omega_N)}^2}{(\omega_q - \omega_N)^2 + \kappa^2} + P_\Delta.
\end{align}

We can see that the radiated power of the elastic scattering corresponds to that of a non-linear driven antenna, and the power due to the emitter fluctuations is given by 
\begin{align}
    P_\Delta = 2\, C_\Delta \left[ \frac{\pi}{2} + \arctan\left(\frac{\omega_q}{\kappa} \right) \right].
\end{align}

The maximal power emitted due to the dipole fluctuations can now be found, and is given by 
\begin{align}
    P_\Delta^{max} \le  \Delta_{\delta d} \frac{ \pi \, g_q^2 }{\kappa} = c \, \Delta_{\delta d} \left(\frac{g_q}{g_0} \right)^2, 
\end{align}
which is obtained in the limit of $\omega_q \gg \kappa$, when the cavity decay rate is much slower than the resonator frequency. We find that the maximal power of the emission originating from the dipole fluctuations depends, obviously, on the strength of the dipole noise $\Delta_{\delta d}$, and more importantly can be tuned by the ratio between the light-matter and resonator-environment coupling.

\section*{\label{sec:conclusion}Conclusions}

The results of this work imply that the non-linear driven resonator, the process of HHG and the non-linear antenna are isomorphic. This is despite their antithetical dynamical properties of being probabilistic and deterministic, respectively. 
This was achieved by using a quantum stochastic analysis of the dynamics in the Markovian regime, and further allowed to obtain the upper bound of the emitted power due to the fluctuations of the non-linear emitter .

The underlying problem of the stochastic noise operators is the origin of the whole discussion of quantum stochastic methods in quantum optics~\cite{gardiner2004quantum}, and considering a time-evolution approach of the corresponding state will lead to stochastic differential equations~\cite{gardiner1985input, gardiner1992wave}. Treating the stochastic increments in the Ito or Stratonovic form within the context of this work will be discussed elsewhere~\footnote{P. Stammer, \textit{et al.}, (in preparation)}, and an analysis via phase-space methods and the Fokker-Planck equation can reveal new insights~\cite{carmichael2013statistical, drummond1981quantum}.
With this work we have therefore paved the way for investigations towards open system descriptions for quantum optical high harmonic generation.

Possibly interesting directions for extending this work are studies of structured environments and quantum noise in HHG, in which one needs to go beyond the quantum regression theorem and resorts to the fluctuation-dissipation theorem~\cite{ford1996there}.
This can open the way for spectroscopy methods using HHG and quantum Langevin approaches~\cite{reitz2025nonlinear}, or even thermodynamic relationships in coherently driven non-linear systems~\cite{schrauwen2025thermodynamic}.
Such studies will further extend the domain of coverage of high harmonic generation. While HHG was first extended from semi-classical towards quantum optics in recent works~\cite{cruz2024quantum, stammer2025theory, gorlach2020quantum, yi2024generation}, this work now bridges the gap to open quantum systems and stochastic quantum optics~\cite{gardiner2004quantum}. 
Ultimately, this work can extend the study of fluctuation-dissipation relations towards ultrafast quantum optics, connects attosecond science with open quantum systems, and can furthermore be used in the context of analog quantum simulation, where an atom in an opto-mechanical cavity allows for the simulation of the \textit{quantum Kramers-Henneberger} system~\cite{arguello2025quantum}.

\begin{acknowledgments}

I would like to express my gratitude to Alexander Carmele for the introduction to stochastic quantum optics.
P.S. acknowledges funding from the European Union’s Horizon 2020 research and innovation programe under the Marie Skłodowska-Curie grant agreement No 847517.
ICFO acknowledges support from: European Research Council AdG NOQIA; MCIN/AEI (PGC2018-0910.13039/501100011033,  CEX2019-000910-S/10.13039/501100011033, Plan National STAMEENA PID2022-139099NB, project funded by MCIN/AEI/10.13039/501100011033 and by the “European Union NextGenerationEU/PRTR" (PRTR-C17.I1), FPI); project funded by the EU Horizon 2020 FET-OPEN OPTOlogic, Grant No 899794, QU-ATTO, 101168628), Fundació Cellex; Fundació Mir-Puig; Generalitat de Catalunya (European Social Fund FEDER and CERCA program.

\end{acknowledgments}

\bibliography{literatur}{}

\appendix
\section*{Appendix}

\subsection{\label{app:dipole_noise}Dipole correlation noise term}

The noise term contributing to the field observables originating from the dipole moment fluctuations is given by 
\begin{align}
    \expval*{\mathfrak{D}^\dagger(t) \mathfrak{D}(t)} = \ &  g_q^2 \ e^{- 2 \kappa t} \int_0^t dt' \int_0^t dt'' e^{-i \omega_q (t' - t'')} e^{\kappa (t'+t'')} \nonumber \\
    & \times \expval{\delta d(t') \delta d(t'')}.
\end{align}

Assuming that the dipole fluctuations are of white noise type with delta-correlated noise structure, and by definition having a vanishing mean $\expval{\delta d(t)} =0$, we consider 
\begin{align}
    \expval*{\delta d(t') \delta d(t'')} \equiv \Delta_{\delta d} \  \delta(t'-t''),
\end{align}
where $\Delta_{\delta d}$ is the magnitude of the dipole correlations, such that 
\begin{align}
    \expval*{\mathfrak{D}^\dagger(t) \mathfrak{D}(t)} \simeq \Delta_{\delta d} \frac{g_q^2 }{2 \kappa} \left[ 1-  e^{- 2 \kappa t} \right].
\end{align}

In the long-time limit, this gives a constant contribution to the field quantities, given by 
\begin{align}
    \lim_{t \to \infty} \expval*{\mathfrak{D}^\dagger(t) \mathfrak{D}(t)} \simeq \Delta_{\delta d} \frac{g_q^2 }{2 \kappa}.
\end{align}

\subsection{\label{app:regression_theorem}Quantum regression theorem}

To use the Onsager-Lax quantum regression theorem (QRT), we shall look at the derivative 
\begin{align}
    \dv{\tau} \expval*{a_q^\dagger (t) a_q(\tau)} = i \expval*{a_q^\dagger(t) [H(\tau), a_q(\tau)]},
\end{align}
where we have used the Heisenberg EOM $\dv{t} A(t) = i [H(t), A(t)]$. 
We can now use that from \eqref{eq:EOM_field} we know how the field operator commutes with the Hamiltonian, such that we have
\begin{align}
\label{app:QRT_ODE}
    \dv{\tau} & \expval*{a_q^\dagger (t) a_q(t+\tau)} =  [- i \omega_q - \kappa] \expval*{a_q^\dagger (t) a_q(t+\tau)} \nonumber \\
    & + g_q^2 \int_0^t dt^\prime e^{(i \omega_q-\kappa)(t-t^\prime)} \expval*{d(t^\prime)d(t+\tau)} ,
\end{align}
where we have used that all the fields are initially in the vacuum.
To proceed, we write the dipole moment correlations as 
\begin{equation}
    \expval*{d(t_1) d(t_2)} = \expval*{d(t_1)} \expval*{d(t_2)} + \expval*{\delta d(t_1) \delta d(t_2)},
\end{equation}
such that the solution to \eqref{app:QRT_ODE} is decomposed in three terms
\begin{align}
    \langle  a_q^\dagger & (t) a_q(t+\tau)\rangle =  e^{- (i \omega_q + \kappa) \tau} \expval*{a_q^\dagger (t) a_q(t)} \nonumber \\
    & + g_q \expval*{a_q^\dagger(t)} \int_0^\tau d\tau^\prime e^{- (i \omega_q + \kappa)(\tau - \tau^\prime)} \expval*{d(t+\tau^\prime)} \\
    & + \Delta_{\delta d} \frac{ g_q^2 }{2 \kappa} e^{- i \omega_q \tau} e^{- \kappa \tau} \left( 1- e^{- 2 \kappa t} \right). \nonumber
\end{align}

In the last term originating from the dipole moment fluctuations, we have again used the delta-correlated noise structure as in End Matter \ref{app:dipole_noise}.

\end{document}